\documentclass[useAMS,usenatbib,referee]{biom}
\usepackage{xcolor}
\usepackage{amssymb}
\usepackage{amsmath}
\usepackage{bigints}
\usepackage{xcolor}
\usepackage{relsize,exscale}
\usepackage[font=normalsize]{subfig}
\usepackage{mwe} 
\usepackage{lscape}
\usepackage{verbatim}
\usepackage{xurl}

\usepackage{subfig}
\usepackage[figuresright]{rotating}

\usepackage{enumitem}
\usepackage[T1]{fontenc}

\title[Bayesian Arc Length Survival Analysis Model (BALSAM)] {Bayesian Arc Length Survival Analysis Model (BALSAM):\\ Theory and Application to an HIV/AIDS Clinical Trial
}

\author
{Yan Gao$^{1,*}$, Rodney A. Sparapani$^{1}$, and Sanjib Basu$^{2}$\\
$^{1}$Division of Biostatistics, Medical College of Wisconsin, Milwaukee, Wisconsin 53226, U.S.A.
\\
$^{2}$Division of Epidemiology and Biostatistics, University of Illinois Chicago, Chicago, Illinois 60612, U.S.A.
\\ $^{*} email:$ yagao@mcw.edu 
}

\begin{document}

\pagerange{\pageref{firstpage}--\pageref{lastpage}} \pubyear{2022}

\volume{XX}
\artmonth{XX}
\doi{XX}


\label{firstpage}


\begin{abstract} 
Stochastic volatility often implies increasing risks that are difficult to capture given the dynamic nature of real-world applications. We propose using arc length, a mathematical concept, to quantify cumulative variations (the total variability over time) to more fully characterize stochastic volatility. The hazard rate, as defined by the Cox proportional hazards model in survival analysis, is assumed to be impacted by the instantaneous value of a longitudinal variable. However, when cumulative variations pose a significant impact on the hazard, this assumption is questionable.  Our proposed Bayesian Arc Length Survival Analysis Model (BALSAM) infuses arc length into a united statistical framework by synthesizing three parallel components (joint models, distributed lag models, and arc length). We illustrate the use of BALSAM in simulation studies and also apply it to an HIV/AIDS clinical trial to assess the impact of cumulative variations of CD4 count (a critical longitudinal biomarker) on mortality while accounting for measurement errors and relevant variables.
\end{abstract}

\begin{keywords}
Cumulative variation; Joint model; Longitudinal biomarker; Stochastic volatility; Survival analysis.
\end{keywords}

\maketitle

\section{Introduction}
\label{s:intro}
Stochastic volatility is ubiquitous in the economy, environment, biology, and health, whether at the population or individual level. Significant biological variations of health-related measurements, physical and mental, are collected on humans and animals, such as blood pressure, heart rate, respiratory rate, brain activity, and mood swings. Many of these measurements are well recognized as biomarkers: a term referring to a broad category of health signs that are objective indications of the internal state, observed externally, which can be measured accurately  (paraphrasing \citet{strimbu:2010}). Current biomarkers present an immense opportunity to accelerate basic science, drug discovery, and medical device development and improve clinical care \citep{Fiore:2011,robb:2016}.

In drug development, validated biomarkers will typically precipitate a shorter study period and smaller sample size in clinical studies, facilitating the efficient development of safe and effective medical products. Traditional clinical trials using the quality or length of life as the primary clinical endpoints will generally have opportunity costs; i.e., they will be more expensive, take longer to complete, and, likely, both will be the case, unnecessarily delaying an effective new drug's approval. Substitution of a short-term, or frequent, biological marker for a rare, or latter, clinical endpoint can substantially reduce sample size and trial duration \citep{lin:1993}. Surrogate endpoints are defined as ``a substitute for a direct measure of how a patient feels, functions, or survives'' \citep{fleming:2012}. Exploring biomarkers to serve surrogate primary endpoints, instead of overall survival (OS),
is becoming increasingly prevalent in clinical trials, especially for cancer and infectious diseases \citep{feigin:2004, farkona:2016, anagnostou:2017,zhang:2020,delgado:2021}. However, the lack of knowledge about the magnitude and duration of a biomarker's effect on a clinically meaningful endpoint, such as OS, compromises the reliability and interpretability of trials designed around a surrogate endpoint.
The guidance of the \citet{FDA:2021} is as follows: ``biomarkers as surrogate endpoints alone do not provide the total picture of benefit and risk of a therapy." These limitations underscore the importance of rigorous evaluation for surrogate endpoints. Moreover, measurement errors often complicate the validity of a biomarker during evaluation. This complex process may have an unwanted impact on the estimation of the actual treatment effect for a clinical outcome. 

The true relationship may be misunderstood when longitudinal biomarkers and survival outcomes are analyzed separately. A better approach would be to propose survival models, such as the Cox proportional hazard model, which treat a longitudinal outcome as a time-dependent covariate while allowing the hazard to vary; yet, this fails to account for measurement errors. As statistical techniques and computing power advance, it is becoming more common to jointly model both types of data simultaneously \citep{Self:1992,Pawitan:1993,De:1994,Tsiatis:1995,FAUCETT:1996,Wulfsohn:1997,Ibrahim:2001,Henderson:2002,Alsefri:2020}.  Longitudinal biomarkers associated with a time-to-event outcome have been jointly analyzed to better evaluate a treatment effect, if any \citep{Ibrahim:2010,Solene:2015,Solene:2017, Guedj:2018}. However, care should be taken to define a survival submodel within a joint model that adequately addresses the two facets of the available measures: (i) the current values as explained by shared random effects, their derivative/slope, or combinations thereof; and (ii) cumulative effects, also known as the total exposure or the area under the curve, which is equivalent to the integral of the longitudinal trajectory from baseline to the current time \citep{Brown:2009}. When the risk of an event depends on the cumulative effect of the time-varying covariate, the current value association structure may not be appropriate given its single time point limitation \citep{Sylvestre:2009}. Meanwhile, distributed lag models are generalized to analyze survival data along with cumulative effects of longitudinal outcomes \citep{Gasparrini:2010,Gasparrini:2014,Gasparrini:2017, Bender:2018,Bender:2019}; these lag models construct a general functional form to represent cumulative effects. 

However, when cumulative variations (defined as the total variability over a longitudinal history) significantly impact the survival outcome, current methods can be incorrect or inefficient. Furthermore, because variations depend on the length of an observation window, existing statistical measurements, including variance, Fano factor, and coefficient of variation \citep{ fano:1947, allison:1978, choloniewski:2020}, cannot adequately capture cumulative variations. Thus, we propose a new measurement defined by arc length, the distance between two points along a parametric curve, to quantify cumulative variations and capture the magnitude and duration of stochastic volatility. Our approach generalizes the $k$-step-ahead sample arc length: a natural measure to quantify volatility in time series \citep{Wick:2014}. Here, we review the background of the relevant methodology from both the Bayesian and frequentist standpoints; however, we intend to take the Bayesian perspective and exploit its many advantages, such as well-known efficient posterior sampling capabilities along with estimation via summaries of the posterior including inherent uncertainty quantification. This re-frames the statistical challenge as quantifying, modeling, and performing inferences concerning cumulative variations in survival analysis. We propose an innovative model that we call the Bayesian Arc Length Survival Analysis Model (BALSAM) to address these issues. Our model accounts for biomarkers' stochastic volatility and measurement errors, along with cumulative variations featured by arc length.

In the bloodstream, circulating CD4 positive T helper cells are a biomarker of the health, or lack thereof, of one's immune system, particularly for those suffering from Human Immunodeficiency Virus (HIV) infection and Acquired Immunodeficiency Syndrome (AIDS). Clinically,
the CD4 count is among the best surrogates of disease progression along with HIV viral load \citep{Tsiatis:1995, Wulfsohn:1997,Dafni:1998}. HIV infection is associated with a progressive increase in viral replication and progressive depletion of CD4 cells 
\citep{Redfield:1988}. We explore this complex relationship between disease prognosis and patient outcomes with a data set from the Community Programs for Clinical Research on HIV/AIDS (CPCRA) randomized clinical trial \citep{Abrams:1994}.
For HIV-infected patients, our analysis aims to assess the impact of CD4 cumulative variation on the hazard rate of mortality, addressed by arc length, while accounting for measurement errors, patient covariates, and experimental treatments.

The rest of the article is organized as follows. First, we detail the model concept and estimation in Section~2, including arc length's definition and computational methods. Next, the results of the simulation studies built upon two illustrative models are provided in Section~3. The statistical results analyzed by BALSAM in the CPCRA trial are provided in Section~4. Finally, concluding remarks and future research are given in Section~5. 

\section{Bayesian Arc Length Survival Analysis Model}
\subsection{Cumulative Variation Measured by Arc Length}
\begin{figure}[ht]
		\centering
    	\includegraphics[width=0.8\textwidth]{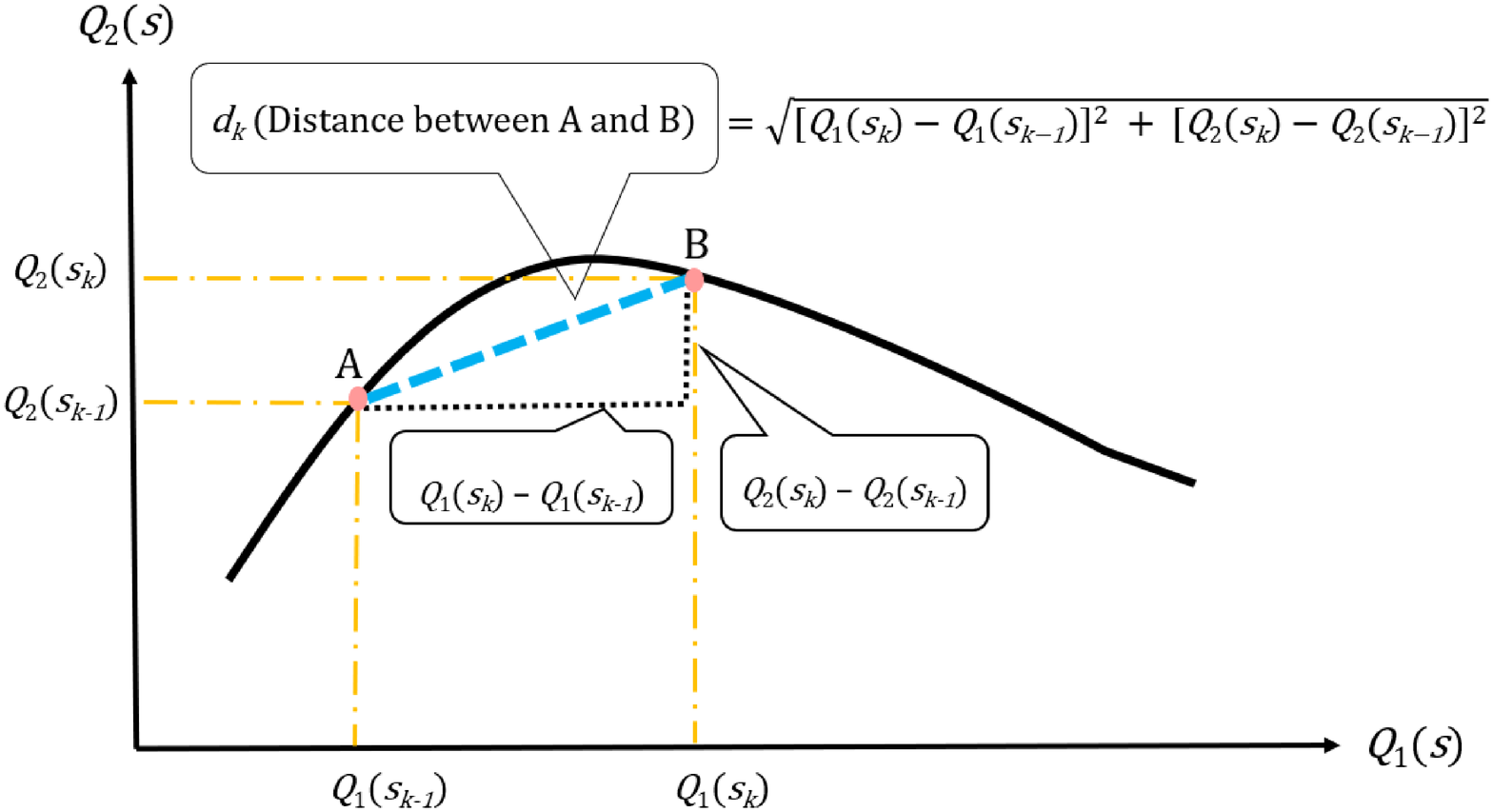} 
		\includegraphics[width=0.73\textwidth]{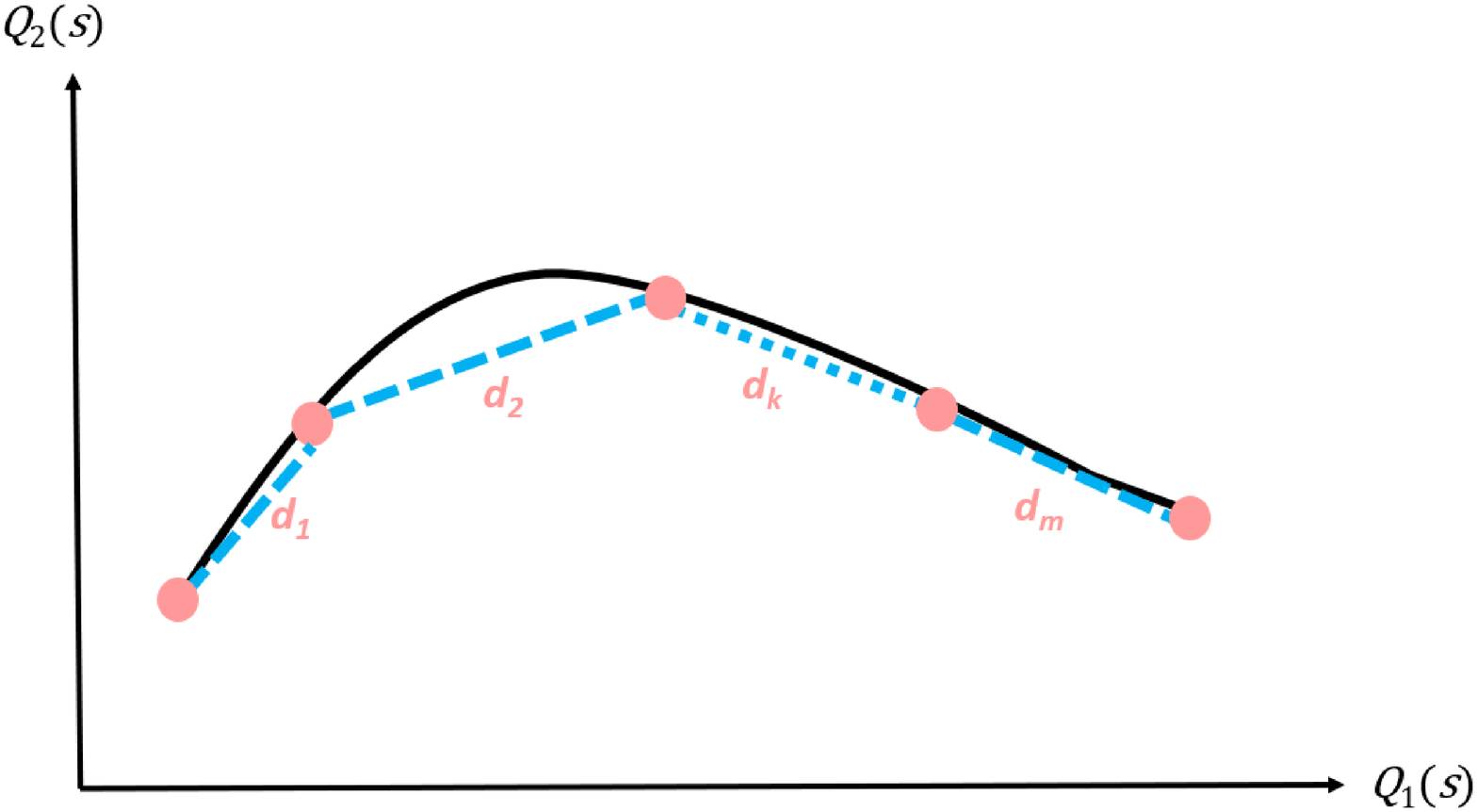} 
		\caption{Arc length of a parametric curve. Both components in $g(s)=( Q_{1}(s), Q_{2}(s))$ are on the two axes, respectively. The arc length between A and B (black solid line) can be approximated by the corresponding segment length (blue dashed line) based on the Pythagorean theorem (top panel). The total approximated distance $L_{m}$ of a parametric curve can be calculated as the sum of $d_{k}$ per partition  (bottom panel). That is, $L_{m}  = d_{1} + d_{2} + \dots + d_{m}, \; \mathrm{where} \; 
d_{k} = d_{k}* \frac{s_{k}-s_{k-1}}{s_{k}-s_{k-1}} = (s_{k}-s_{k-1})\sqrt{\big[\frac{Q_{1}(s_{k}) -Q_1(s_{k-1}) } {s_{k}-s_{k-1}}\big ]^{2} + \big[\frac{ Q_2(s_{k}) -Q_2(s_{k-1}) } {s_{k}-s_{k-1}} \big]^{2}} = (s_{k}-s_{k-1})\sqrt{Q_{1}'(\vartheta_{k})^{2} + Q_{2}'(\varphi_{k})^{2}}$. The values $\vartheta_{k}$ and $\varphi_{k}$ are guaranteed by the mean value theorem and must be in $[s_{k-1},s_{k}]$. As the mesh size of the partition goes to zero,  $L_{m}$ must converge to the arc length $L=\int \sqrt{ Q_{1}'(s)^{2} + Q_{2}'(s)^{2} } \; \mathrm{d} s $.}
	 \label{arc}
\end{figure}

 Define a two-dimensional variation function $g(s)=\big(Q_1(s), Q_2(s))$ that is a parameterized curve where (a) $g : [0, \; t] \rightarrow \mathbb{R}^{2}$ is a continuously differentiable vector function; and (b) $Q_1'(s)$ and $Q_2'(s)$ are the corresponding continuous derivatives for the curve that is traversed once for $s \in [0, \; t]$. If we define the distance norm as $|g'(s)| = \sqrt{Q_1'(s)^{2}+Q_2'(s)^{2}}$, then the cumulative variation $G(t)$, or arc length, is the distance between two points along this function as computed by the following integral \citep{Briggs:2015}.
\begin{equation} \label{arc_len}
G(t) = \int_{0}^{t} |g'(s)| \; \mathrm{d}s
\end{equation}
While we restrict our attention to two dimensions, note that $g(s)$ is easily extended to a $d$~dimensional vector function like so: $g(s)=(Q_{1}(s), \; Q_{2}(s), \; \ldots, \; Q_{d}(s) )$. 
Figure \ref{arc} displays how to utilize the segment between two points, A and B, to approximate the arc length in a plane. As the points get close, the segment approaches the curve. 

Because the measurement trajectory evolves as a function of time, from here on, we use the specific definition $g(s) = \left(s,Q(s) \right) $ aligning with a longitudinal data structure. 
Therefore, $G(t)$ can be imagined as the length of the curve parameterized by the time $s$ and the longitudinal covariate $Q(s)$: $G(t) = \int_0^t 
\sqrt{1+Q'(s)^2} \mathrm{d}s$.
In many cases, there are no closed-form solutions for arc length and numerical integration becomes necessary. Here, we contrast two methods of numerical integration that might be considered. Although Romberg integration does not require the form of the derivative, $g'(s)$; however, Romberg integration may be very computationally demanding. The Trapezoidal rule requires that the derivative $g'(s)$ has to be specified and it is computationally friendly whenever derivatives are readily available. Provided that $|g'(s)|$ is integrable, the Trapezoid rule approximation, while using $m$ equally spaced sub-intervals on $[0, \; t]$, is expressed as follows: 
\[G(t) \approx \left( \frac{1}{2} |g'(0)| + \sum_{k=1}^{m-1} |g'(s_{k})| +  \frac{1}{2} |g'(t)| \right) \Delta s \]
where $\Delta s =t/m$ and $s_{k}=k \Delta s$ for $k=1, \ldots , m-1$.
\subsection{Model and Notation}
\label{Nota_data}
For each subject, the observed survival time is denoted by $t_{i} = \min(c_{i},\; y_{i})$, where $y_{i}$ is the event time, $c_{i}$ is the censoring time,
and $i=1, \dots, n$ indexes the subjects. The event indicator is $\delta_{i}=I(y_{i} \leq c_{i})$ where $I(.)$ is the predicate indicator function: $I(\mbox{true})=1$ and $I(\mbox{false})=0$. No cumulative variations are assumed for the
survival covariates denoted by the vector $\bm{x}_i=\{x_{i1}, \dots, x_{iP}\}$. 
Consider a single longitudinal variable  for subject $i$ measured at time points $\bm{s}_{i} =\{s_{ij}\}$
where $j=1,\dots, n_{i}$; and these collected measurements with error are denoted as $\bm{z}_{i}=\{z_{ij}\}$
along with the corresponding unobserved true values ${Q}_i(s)=\{Q_{ij}\}$. Thus, the observed data for each subject is denoted by ($t_{i}, \delta_{i}, \bm{s}_{i}, \bm{x}_{i}, \bm{z}_{i}$). The cumulative variation $G_{i}(t) = \int_{0}^{t_{i}} \sqrt{1+ Q_{i}'(s)^{2}} \; \mathrm{d}s$ is the total variability of a longitudinal outcome over the survival history. Conditionally independent right-censoring is assumed. Basically, BALSAM consists of two joint submodels. 
\begin{align} \label{eq:M_0}
 \mbox{Survival model}:\qquad & & \lambda_i(t)\quad & = \lambda_{0}(t) 
 \exp \left\{\bm{x}_{i}'\bm{\beta}+\alpha G_{i}(t) \right\} \\
\mbox{Longitudinal model}: & &
E[z_{i}(t)] & = \eta^{-1}(Q_{i}(t)) \mbox{\ where\ } \eta(.) \mbox{\ is a link function} \nonumber
\end{align}
The mathematical concept of arc length (\ref{arc_len}) is extended via  BALSAM to three statistical foci: (i) a joint survival model; (ii) the Bayesian paradigm; (iii) distributed lag models. In other words, we add a new element, arc length, into the intersection rather than the union of these components. This new statistical fusion will transform the current parallel development of each element into a unified more powerful triad. To the best of our knowledge, we are the first to propose this novel idea. 

The goal is to achieve high accuracy in dynamic hazard prediction while accounting for measurement errors and cumulative variations of a longitudinal outcome. We modify the original Cox model formulation for survival data, where the hazard $\lambda_{i}(t)$ depends on a longitudinal outcome through its cumulative variation, whose association degree is measured by the parameter $\alpha$. The generalized mixed effect model is applied to analyze the longitudinal outcome $z_{i}(t)$, which is subject to measurement errors. $Q_{i}(t)$ is assumed to be a smooth function that polynomial splines can approximate. Although actual data is collected at only a finite set of discrete time points; the hazard, $\lambda_i(s)$, as well as the measurement process $Q_i(s)$ and the arc length $G_i(s)$ are defined at all time points $s \in [0, t_i]$. This is an important consideration, particularly for $Q_i(s)$, since we do not have to make precarious restrictive assumptions like last value carried forward for the measurements~$z_i(s)$.
\subsection{Statistical Estimation}
The Bayesian paradigm is adopted for statistical estimation. The MCMC algorithms guarantee convergence to the posterior distributions with minimal requirements on the target. Yet, as the number of iterations grows with the dimension of the problem and the volume of the data, random walk posterior sampling algorithms, including Gibbs sampling and Metropolis–Hastings, may induce an extremely lengthy exploration because they are not aware of the global support of the distribution \citep{Robert:2018}. 
One way to address this issue is to decrease the step size of the random walk to avoid frequently bypassing the highest probability region. However, this will often lead to highly auto-correlated samples and lengthy iterations due to the many more small steps required to achieve sufficient accuracy. Hamiltonian Monte Carlo (HMC), particularly the No-U-Turn Sampler (NUTS), are more recent methods that adopt Hamiltonian dynamics concepts to reduce this auto-correlation and restrains the posterior samples to non-negligible regions  \citep{Duane:1987, Neal:1993, Neal:1996, MacKay:2003,Hoffman:2014,Robert:2018}. Methods built upon the Hamiltonian dynamics usually converge faster, particularly for complicated models in a high-dimensional parameter space, such as joint models. 

In Gelfand-Smith Bayesian bracket notation \citep{gelfand:1990}, let $[\bm{\theta}]$ and $[\bm{y}|{\bm\theta}]$ denote the prior distribution and likelihood, respectively, e.g., 
if $\bm{y}|\bm{\theta}\sim f(\bm{y}|\bm{\theta})$, then $[\bm{y}|{\bm\theta}]=f(\bm{y}|\bm{\theta})$ where $f(.)$ is the density function. Under conditional independence assumption, the observed joint likelihood for each subject can be decomposed into three components: survival likelihood, longitudinal likelihood and random effects likelihood. Let $\bm{\theta}=( \bm{\theta}_{S},\bm{\theta}_{L}, \bm{\theta}_{R})$ represent all parameters, where $\bm{\theta}_{S}$, $\bm{\theta}_{L}$ and $\bm{\theta}_{R}$ denote the parameters associated with survival, longitudinal, and random effects parameters, respectively. To simplify the notations, the conditional variables are omitted in the posterior distribution as expressed below:
\begin{align}\label{long_eqn}
&[\bm{\theta}|\bm{y}] \propto 
[\bm{\theta}]  \prod_{i=1}^{n} \left\{ \int_{-\infty}^{\infty} [t_{i},\delta_{i}|\bm{\theta}_{S}] \; [{\bm{z}_{i}}|\bm{\theta}_{L}] \; [\bm{b}_{i}|\bm{\theta}_{R}] \; \mathrm{d} \bm{b}_{i} \right\}   
& 
[t_{i},\delta_{i}|\bm{\theta}_{S}] & 
\propto \lambda_i(t)^{\delta_{i}}  S_{i} (t) \\
&[  {\bm{z}_{i}}|\bm{\theta}_{L}] \propto
\exp \sum_{j=1}^{n_{i}} \left\{{ \frac{z_{ij}\bm{\theta}_{L}-a(\bm{\theta}_{L})} {\phi(\nu)} +q(z_{ij},\nu) } \right\} 
& 
\bm{b}_{i}|\bm{\theta}_{R} & \sim f_{{b}}(\bm{b}_i) \nonumber
\end{align}
where we leave $ f_{{b}}(.) $
unspecified for the moment. An exponential family \citep{Mccullagh:1989, McCulloch:2003} is assumed for $z_{ij}$ in (\ref{long_eqn}), 
where $\bm{\theta}_{L}$ and $\nu$ represent the canonical and dispersion parameters, respectively. These known functions $\phi(\cdot),\; a(\cdot), \mathrm{and} \; q(\cdot)$ vary from one exponential family to another. In terms of modeling, either parametric, or nonparametric, regression methods could be employed.
\subsection{Illustrative Models}
We consider two models for illustration. Model I demonstrates the computational convenience when the closed-form solution is available. For simplicity, the constant baseline hazard is assumed: $\lambda_{0}(t)=\lambda$. A single time-independent covariate $x_{i1}$ is considered for the hazard modeling. Assume the random intercept and slope effects have a bivariate normal distribution in the longitudinal submodel, as follows.
\begin{eqnarray*}
\begin{pmatrix}b_{i0}\\ b_{i1}\\
\end{pmatrix} & \sim & N_{2}\left[ \bm{\mu}  =  \left ( \begin{array}{c}
\mu_{1}\\
\mu_{2}\\
\end{array}\right), \; \bm{\Sigma}  = \left(\begin{array}{ccc}
\sigma_{1}^2 & \sigma_{21}\\
\sigma_{21} & \sigma_{2}^2\\
\end{array}\right) \right]
\end{eqnarray*}
Therefore, the cumulative variation is $G_{i}(t) =  \bigintss_{\; \; 0}^{t_{i}} \sqrt{1+b_{i1}^{2}} \; \mathrm{d} s  = t_{i} \sqrt{1+b_{i1}^{2}}$. 
Model I is given by 
\begin{equation} \label{eqn:M1}
\begin{array}{l}
 \lambda_{i}(t) =\lambda \exp\left\{x_{i1}\beta + \alpha t_{i} \sqrt{1+ b_{i1}^{2}}\right\} \\
z_{ij}=b_{i0}+b_{i1} s_{ij} + \varepsilon_{ij} 
\end{array}
\end{equation}
where $\varepsilon_{ij} \sim N(0, \,\sigma^{2})$.
\noindent The popular random walk method, Gibbs sampling, will be utilized for posterior estimation in Model I given the simple calculation of arc length. 

Model II replaces the linear regression in Model I with nonparametric regression for the longitudinal data to capture subject heterogeneity. Bayesian nonparametric modeling combines the flexibility of nonparametric models with the exact inference provided by the Bayesian inferential machinery \citep{berry:2002,ruppert:2003, crainiceanu:2005}. Given $\bm{b}_{i} \sim  N_{K} \left ( \bm{\mu}, \; \bm{\Sigma} \right )$, $Q$ is assumed as a smooth function which can be well approximated by piece-wise polynomial splines. The popular B-spline basis functions denoted as $B(.)$, have minimal support with respect to a given degree, smoothness, and domain partition. Model II is expressed as follows. 
\begin{align} \label{eqn:M2}
& \lambda_{i}(t) =\lambda  \exp \left \{ x_{i1} \beta + \alpha G_{i}(t)  \right \} & \; \mathrm{where} \; G_{i}(t) = \bigints_{\,0}^{t_{i}} \sqrt{1+ \left \{ \sum_{l=1}^{K} b_{il}B'_{l}(s)
\right \}^{2}} \; \mathrm{d} s 
\\
& z_{ij} =Q_{ij}+\varepsilon_{ij}= 
\sum_{l=1}^{K} b_{il}B_{l}(s_{ij}) + \varepsilon_{ij} &   \nonumber
\end{align}


The popular accelerating MCMC method, NUTS, will be utilized for posterior estimation in Model II. Its advantages originate from the automatic tuning of two critical parameters in HMC. 
Thus, the hand-tuning process, often a drawback of HMC, can be avoided by NUTS 
\citep{Andrieu:2008, Nesterov:2009}. 

\section{Simulation Study} \label{sim_stu}
The simulation method developed by \citet{Bender:2005} for survival data is modified to serve BALSAM.
Let the event time $t$ be a random variable with the cumulative distribution function $F$, then $u = F(s) = P(t<s)$ follows a uniform distribution on the interval $[0, 1]$ denoted as $u \sim \; \mathcal{U}_{[0,1]}$. Similarly, if $u \sim \; \mathcal{U}_{[0,1]}$
and $v=1-u$, then $v \sim \; \mathcal{U}_{[0,1]}$ (N.B. the survival function is  $S(s)=P(t>s)=1-F(s)=\exp(-H(s))$
where $\lambda(t)$ and $ H(s) $ are the hazard and cumulative hazard functions respectively: 
$H(s)=\int_0^s \lambda(t) \mathrm{d} t\;$). 
The probability integral transformation states that if $u$ has a uniform distribution on $[0, 1]$ and $t$ has a cumulative distribution $F(s)$, then the random variable $F^{{-1}}(u)$ has the same distribution as $t$. Let $v = \exp \{-H(t) \}  \sim \mathcal{U}_{[0,1]}$, thus:
    $t=H^{-1}\left(-\log(v)\right)$ where 
$H^{-1}(.)$ is the inverse function of $H(.)$.  It is convenient if a closed-form expression exists for $H^{-1}$(.); otherwise, the inversion problem is transformed into the unique root-finding problem since $H(.)$ is monotonically increasing. 

Based on Model I (\ref{eqn:M1}), we conduct a simulation study with 1,000 simulated data sets. The Markov chain sampling method is used for the posterior analysis with the following MCMC settings: discarding 5,000 burn-in samples and keeping 20,000 iterations with a thinning interval of 5 (keeping 1 out of every 5). As shown in Table \ref{Model12}, the 95\% coverage rate from the posterior results is approximately attained for all parameters, i.e., all parameters within 94\% to 99\% coverage. 
It suggests that overall, Model I fits the data well.

\begin{table}[ht]
\subfloat[Simulation Results for Model I] {

\begin{tabular*}{\textwidth}{@{\extracolsep{\fill}}lll} 
 \hline 
Model I (\ref{eqn:M1}) Parameters & Data-generating Value & 95\% Coverage Rate\\ 
 \hline 
($\lambda$,\;  $\beta $,\;  $\alpha $, \; $\sigma^2$ )
& (0.02, \;0.05,\; 0.25,\; 4.00)
& (0.99,\; 0.95,\; 0.97,\; 0.97)\\ 
$\bm{\mu}$\;=\;( $\mu_{1}$,\; $\mu_{2}$) & (1.20,\;0.25) & (0.97,\; 0.96)\\ %
$\bm{\Sigma}$\;=\;
$\begin{pmatrix}
\sigma_{1}^2 & \\
\sigma_{12} & \sigma_{2}^2
\end{pmatrix}$
 &
$\begin{pmatrix}
2.00 & \\
3.00 & 5.00
\end{pmatrix}$
&
$\begin{pmatrix}
 0.95 & \\
 0.94 & 0.95
\end{pmatrix}$ \\
 \hline
\end{tabular*} } 
\vspace{0.5cm}

\subfloat[Simulation Results for Model II ] {
\begin{tabular*}{\textwidth}{@{\extracolsep{\fill}}lll}
\hline 
Model II (\ref{eqn:M2}) Parameters & Data-generating Value & 95\% Coverage Rate\\ %
\hline 
($\lambda$, \; $\beta$,\;$\alpha$,\;$\sigma^2$)
& (0.02,\;0.05,\;0.25,\;4.00)
& (1.00,\;0.98,\;0.94,\;0.93)\\ 
$\bm{\mu}$\;=\;($\mu_{1}$,\;$\mu_{2}$,\;$\mu_{3}$,\;$\mu_{4}$ )
& (1.20,\;0.25,\;1.20,\;0.25) & (0.98,\; 0.96,\;0.94,\;0.93)\\ %
$\bm{\Sigma}$\;=\;
$\begin{pmatrix}
\sigma_{1}^2 &  & & \\
\sigma_{21}  & \sigma_{2}^2 & & \\
\sigma_{31} & \sigma_{32} & \sigma_{3}^2 & \\
\sigma_{41} & \sigma_{42} & \sigma_{43} & \sigma_{4}^2 
\end{pmatrix}$
& 
$\begin{pmatrix}
6.60 &  & &\\
5.24 &7.85 & &\\
5.24& 5.80 &6.53&\\
5.93& 6.08 &4.57&6.43
\end{pmatrix}$
  &
$\begin{pmatrix}
0.93 &  & &\\
0.91 &0.96 & &\\
0.95& 0.92 &0.94&\\
0.97& 1.00 &0.93&0.96
\end{pmatrix}$\\

 \hline
\end{tabular*} }

\vspace{1cm}
 \caption{Performance of BALSAM in simulation studies. For model I (\ref{eqn:M1}), 1,000 simulated data sets are generated. For Model II (\ref{eqn:M2}), 97 out of 100 data sets are generated due to computational issues in the other 3 data sets. For each simulated data set, the binary variable \textit{cover} = 1 when the 95\% credible interval of a parameter in the posterior distribution includes the data-generating value; otherwise, \textit{cover} = 0. The 95\% coverage rate is defined as the mean of \textit{cover} over all simulated data sets. Only the lower triangular components of the variance matrix $\bm{\Sigma}$ are presented, given its symmetric property.}
 \label{Model12}
 
\end{table}

Model II (\ref{eqn:M2}) is used to analyze 97 simulated data sets (we planned for 100, but 3 data sets had computational issues). This nonparametric model consists of four B-spline basis functions: $\{B_{l}\}^{4}_{l=1}$. There are three knots, including one inner knot and two boundary knots. The order of the B-splines is fixed at 3. The MCMC settings for NUTS are as follows: discarding a 1,000 burn-in samples, keeping 2,000 sampling iterations with no thinning. The 95\% coverage rates have been approximately attained for all parameters in the range 91\% to 100\% (see Table \ref{Model12}). 
These results are very promising especially given the error due to the approximation of arc length since the closed-form solution of the integration is not feasible in Model II. 

\section{Application in HIV/AIDS Clinical Trial} \label{Appl}
Human CD4 cells play a pivotal role in immunology and immunotherapy 
\citep{zanetti:2015,Borst:2018,derogatis:2021,Gonza:2021}. A CD4 count, with a normal range of 500 to 1500  cells/mm${^3}$, is a common laboratory test to assess the viability of the immune system by measuring the number of CD4 positive T helper cells. This well-known biomarker plays a crucial role in diagnosing and treating infectious diseases, including HIV/AIDS \citep{lin:1993,fleming:2012,Awoke:2019, Battistini:2021}. One indication of an AIDS diagnosis is when the CD4 count drops below 200 cells/mm${^3}$. The CPCRA study was a multicenter, randomized, open-label, community-based clinical trial. It aimed to compare the efficacy and safety of two antiretroviral drugs: didanosine (ddI) vs. zalcitabine (ddC), in treating HIV patients for whom the standard zidovudine (AZT) therapy did not succeed \citep{Abrams:1994}. All three drugs are nucleoside reverse transcriptase inhibitors that prevent reverse transcription of the HIV genome, thereby inhibiting viral replication. The first treatment for HIV, AZT, was first described in 1964 and approved in the United States in 1987 \citep{reeves:2007}. The drug ddI was first described in 1975 and approved in the United States in 1991. The FDA approved the third antiretroviral ddC in 1992 as a monotherapy and in 1996 for use in combination with AZT. Our analysis aims to assess the impact of CD4 count cumulative variation on the hazard rate while accounting for measurement errors and treatment effects. Our approach may help advance the current evaluation of the CD4 biomarker as a surrogate endpoint in immunotherapy.

\begin{sidewaysfigure}[ht]
\centering
        \includegraphics[height=34em,width=1\linewidth]{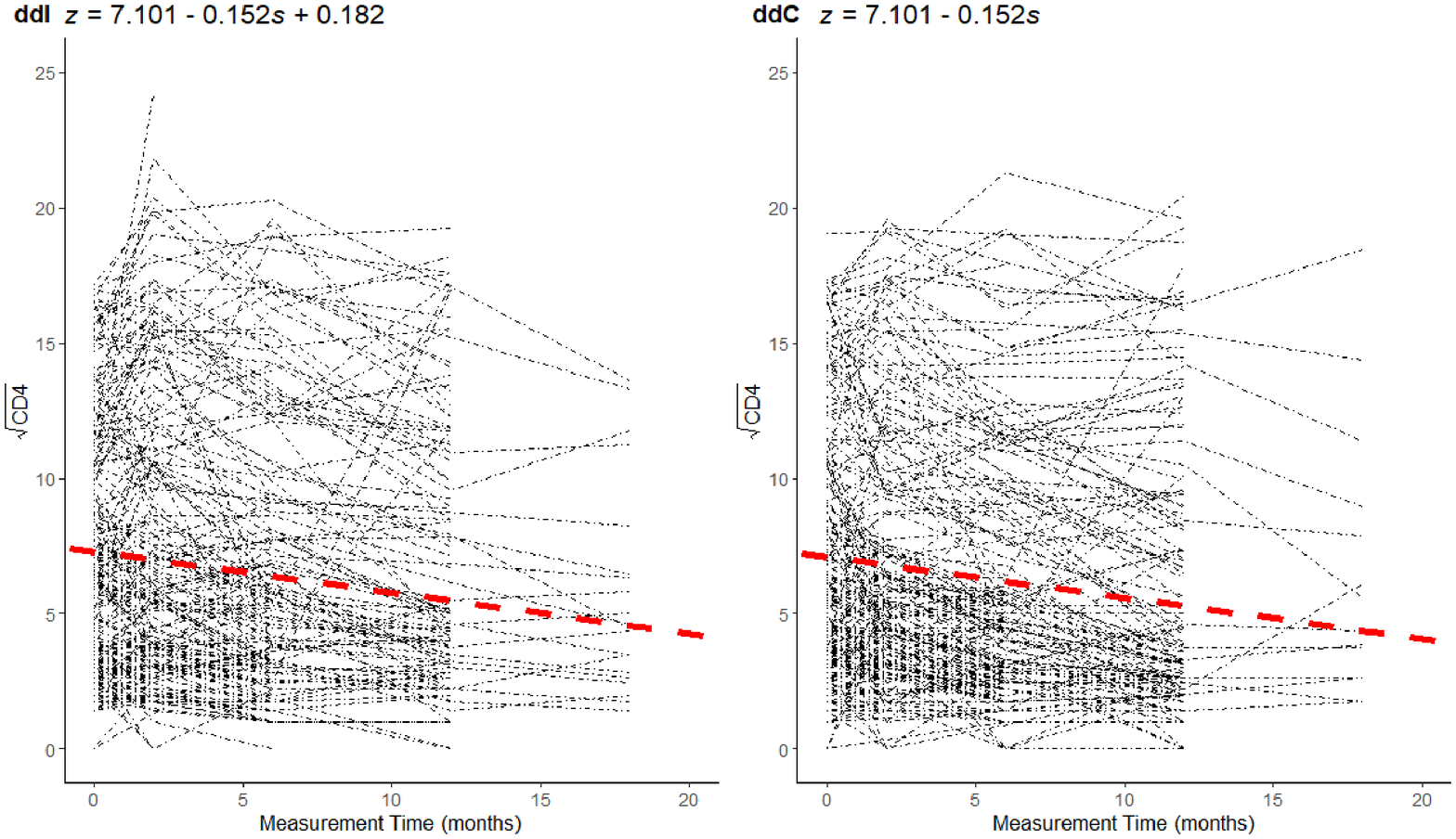}
   \caption{Longitudinal data visualization in the HIV/AIDS clinical trial. The square root of CD4 counts is on the vertical axis and the measurement time (in months) is on the horizontal axis in the spaghetti plot for the two treatment groups (ddI vs. ddC). The fitting curve from BALSAM (red dashed line) is drawn according to the population-level estimates, that is, the posterior mean of the parameters, as expressed in the formula on the top of each treatment group.}
\label{study2:long}
\end{sidewaysfigure}

A total of 467 HIV-infected patients were enrolled from December 1990 through September 1991. Using a permuted-block design, 230 and 237 patients were randomized to two treatment groups, ddI and ddC, respectively. This study measured CD4 count at baseline, 2, 6, 12, and 18 months. There are 1405 observations and nine variables in the analysis data set. One important variable, Karnofsky score, in the analysis by \citet{Abrams:1994}, was not available for our re-analysis. The total number of deaths is 188. The square root transformation is applied to CD4 count, $\sqrt{CD4}$, so that its distribution is more symmetric. Figure \ref{study2:long} suggests that overall $\sqrt{CD4}$ displays a similar trend between treatment groups while demonstrating certain variations within, and between, patients. The Kaplan-Meier method estimates that the median survival time in the ddC group is about 19.1 months. The median survival time is not attained for ddI due to data limitations. A Cox proportional hazards analysis reports a hazard ratio of ddI over ddC is around 1.23. The 95\% CI of the hazard ratio covers 1, suggesting the similar efficacy of these two treatments in prolonging patients' lives. However, the Cox model may not address the following research questions well. How to account for the measurement errors of CD4 count in the survival analysis? What is the association of CD4 cumulative variations with OS independent of the treatment effect while evaluating CD4 count as a surrogate endpoint? 

We reanalyzed these data using Model Ia (\ref{eqn:M1_a}) to examine whether the CD4 cumulative variation could impact OS while adjusting for the following binary variables: treatment (ddI vs. ddC), sex (male vs. female), response to zidovudine (treatment failure vs. intolerance), and previous AIDS-defining condition at study entry (yes vs. no). For each subject, these variables are denoted by $x_{i1}$ to $x_{i4}$, consisting of the four-dimensional vector $\bm{x}_{i}$. For each variable, the latter is the reference group. The number of patients is 61, 91, 122, 169, and 24 for 1, 2, 3, 4, and 5 CD4 measurements, respectively. More than half of patients had less than, or equal to, 3 measurements. Figure \ref{study2:long} suggests $\sqrt{CD4}$ demonstrates a linear trend over time; higher-order trends are almost negligible for the majority of patients. Thus, the random intercept and random slope were considered for the longitudinal modeling. 
\begin{equation} \label{eqn:M1_a}
\begin{array}{l}
 \lambda_i(t_{i}) =\lambda \exp\bigg\{\bm{x}_{i}'\bm{\beta} + \alpha t_{i} \sqrt{1+ b_{i1}^{2}}\bigg\}
\\
z_{ij}=b_{i0}+b_{i1} s_{ij} + \gamma x_{i1}  + \varepsilon_{ij} 
\end{array}
\end{equation}

\begin{sidewaysfigure}[ht]
\centering
   \includegraphics[height=26em,width=1\linewidth]{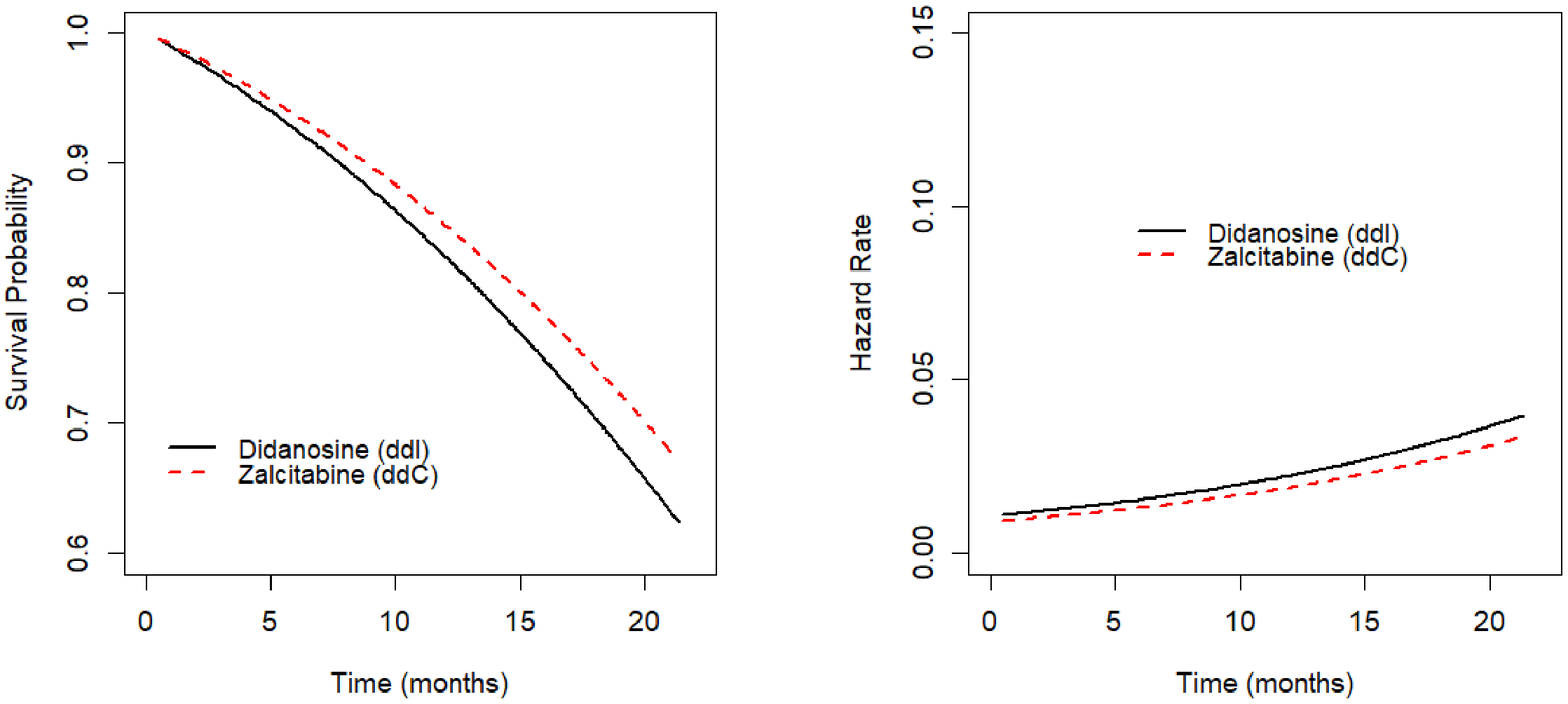}
   \caption{Survival data fitting curves in the HIV/AIDS clinical trial. The estimated survival probability by BALSAM is on the vertical axis, and the time to death is on the horizontal axis (left panel). The estimated hazard rate by BALSAM is on the vertical axis (right panel). The fitting curve is drawn according to the posterior means. The fitted cumulative variation is calculated based on the closed-form formula: $G(t)=t\sqrt{1+\mu_{1}^2}$. All covariates (sex, response to zidovudine, and previous AIDS-defining condition) except treatment are fixed at the reference group coded by 0 in the model.}
\label{study2:surv}
\end{sidewaysfigure}

\begin{sidewaysfigure}[ht]
\centering
   \includegraphics[height=38em,width=0.8\linewidth]{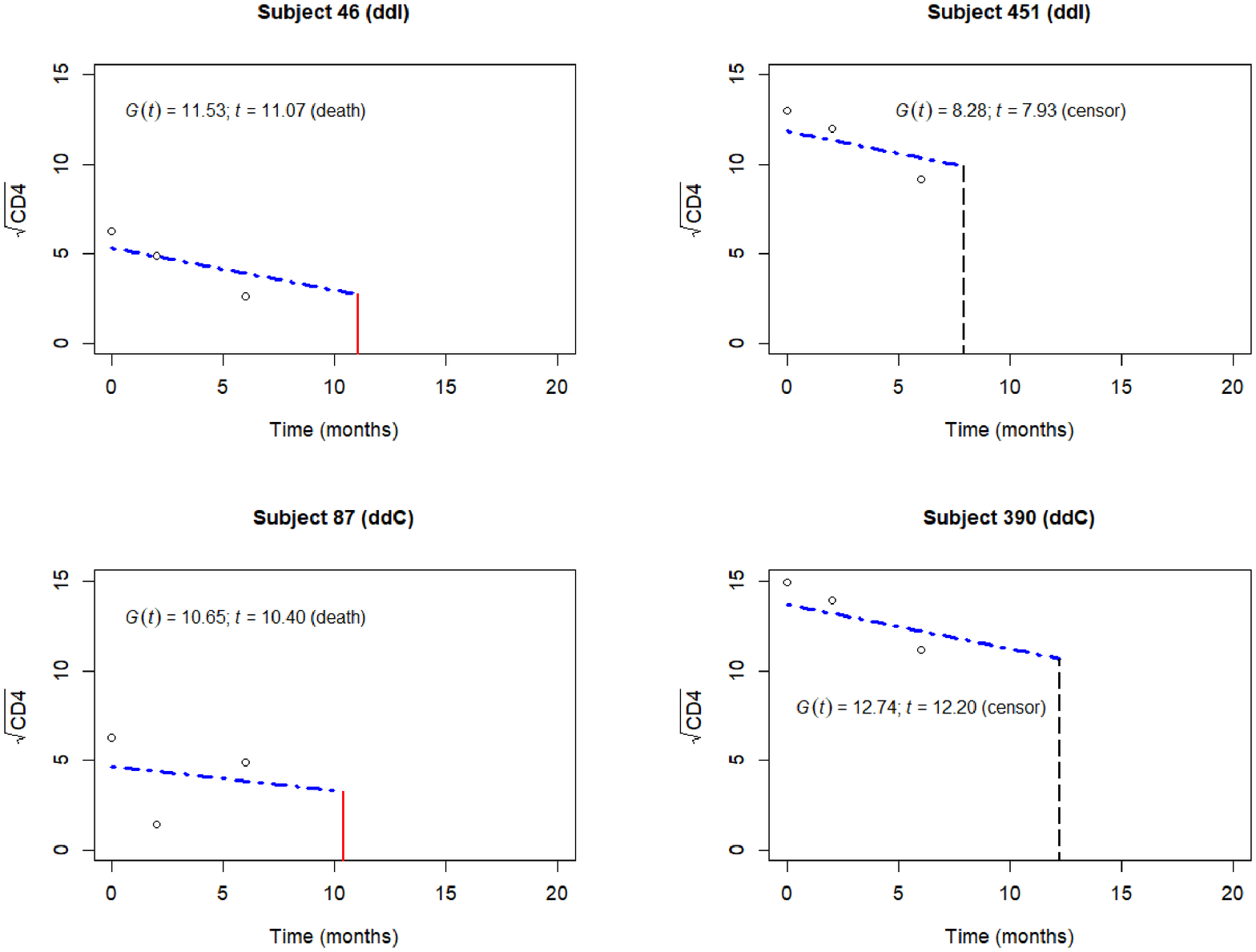}
   \caption{CD4 fitting curves for four patients in the HIV/AIDS clinical trial. The estimated cumulative variation and observed survival time are denoted by $G(t)$ and $t$, respectively. The vertical bar represents the time equal to $t$ (death: red solid line; censor: black dashed line). The blue dashed line is the fitting curve for the individual patient, where different points represent the observed $\sqrt{CD4}$ values over time. The length of the blue line up to the vertical bar is the value of $G_{i}(t)=t_{i} \sqrt{1+b_{i1}^{2}}$.}
   \label{four_pats}
\end{sidewaysfigure}

The survival fitting curves based on (\ref{eqn:M1_a}) are provided in Figure \ref{study2:surv}. The individual fitting curves for four patients are displayed in Figure \ref{four_pats} as an illustrative example. Table \ref{CD4} presents the posterior results analyzed by BALSAM. Successful MCMC convergence is guaranteed by the result that all $\hat{R}$ values $<$ 1.01. The 95\% credible interval of $\beta_{1}$, [-0.093, 0.425], and the posterior probability $P(\beta_{1}>0|\bm{y}) \approx 0.90$ suggests that the two treatments did not demonstrate a significant difference in mortality while adjusting for the other covariates, which is consistent with the results in the Cox model. The posterior mean of $\alpha$, 0.061, and the posterior probability $P(\alpha>0|\bm{y}) \approx 1.0$ depict strong statistical evidence of an association between the $\sqrt{CD4}$ cumulative variation and the hazard. The positive posterior mean and the 95\% credible interval suggest that a greater $\sqrt{CD4}$ cumulative variation could increase the hazard when adjusting for covariates. The estimated posterior mean of the overall intercept of $\sqrt{CD4}$, $\mu_{1}$, is 7.101. In the CD4 scale, this value translates to $7.101^{2}$ $\approx$ 50.42. Thus, the enrolled HIV patients had much lower values of CD4 counts than the normal range. The estimated posterior mean of the overall slope, $\mu_{2}$, is -0.152, coupled with a credible interval [-0.184, -0.120]. This analysis utilizes a linear form to model the longitudinal trajectory of $\sqrt{CD4}$. When the slope is negative, it suggests the overall trend of the CD4 count decreases over time. The 95\% credible interval of $\gamma$, [-0.324, 0.696], and the posterior probability $P(\gamma>0|\bm{y}) \approx 0.76$ suggest the two treatments did not significantly differ in impacting CD4 count. The cumulative variation can be interpreted as the magnitude of CD4 variations here, suggesting a greater CD4 decrease could increase the hazard after adjustment. 

\begin{table}[ht]
\centering
\begin{tabular}{lrrrrr}
\hline
Model Ia (\ref{eqn:M1_a}) Parameters  & Mean (SD) & 2.5\% &97.5\% & Rhat & ESS\\ 
\hline
\multicolumn{6}{l} {Survival Submodel}\\
\hspace{4mm} Constant baseline hazard ($\lambda$) &  0.009 (0.003) & 0.005 & 0.016 & 1.000 & 17184\\ 
\hspace{4mm} Coefficient of treatment ($\beta_{1}$) &  0.166 (0.133) & -0.093 & 0.425 & 1.000 & 27000\\ 
\hspace{4mm} Coefficient of sex ($\beta_{2}$) &  -0.376 (0.242) & -0.830 & 0.117 & 1.000 & 11541\\ 
\hspace{4mm} Coefficient of AZT ($\beta_{3}$) &  0.157 (0.164) & -0.163 & 0.480 & 1.000 & 14917\\ 
\hspace{4mm} Coefficient of PreAIDS ($\beta_{4}$) &  1.283 (0.225) & 0.855 & 1.734 & 1.000 & 15816\\ 
\\
\multicolumn{6}{l} {Cumulative Variation }
\\
\hspace{4mm} Association coefficient ($\alpha$) &  0.061 (0.015) & 0.033 & 0.090 & 1.000 & 27000\\ 
\\
\multicolumn{6}{l} {Longitudinal Submodel}
\\
\hspace{4mm} Overall mean ($\mu_{1}$) & 7.101 (0.256) & 6.595 & 7.599 & 1.000 & 5362\\ 
\hspace{4mm} Overall slope ($\mu_{2}$) &  -0.152 (0.016) & -0.184 & -0.120 & 1.000 &  7367\\
\hspace{4mm} Coefficient of treatment ($\gamma$) & 0.182 (0.259) & -0.324 & 0.696 & 1.001 & 4232\\
\hspace{4mm} Variance of random error ($\sigma^2$) & 3.016 (0.164) & 2.712 & 3.351 & 1.000 & 27000\\ 
\hspace{4mm} Variance of random intercept ($\sigma_{1}^{2}$) & 21.096 (1.520) & 18.300 & 24.275 & 1.000 & 9402\\ 
\hspace{4mm} Covariance ($\sigma_{21}$) & -0.129 (0.075) & -0.280 & 0.015 & 1.000  & 17103\\ 
\hspace{4mm} Variance of random slope ($\sigma_{2}^{2}$) & 0.037 (0.006) & 0.026 &0.050 & 1.000 & 22147\\ 
\hline
\end{tabular}
\vspace{1cm}
 \caption{BALSAM analysis results for the HIV/AIDS clinical trial. The posterior results analyzed by Model Ia (\ref{eqn:M1_a}) is based on the Gibbs sampling method. There are totally 3 Markov chains, along with 100,000 iterations, 10,000 burn-in and 10 thin rate per chain. The 95\% credible interval covers [2.5\%, 97.5\%]. ESS represents Effective Sample Size. AZT denotes response to zidovudine. PreAIDS denotes previous AIDS-defining condition at study entry. The deviance information criterion (DIC), a goodness-fit criterion, is 93408640, estimating the expected predictive error. The effective number of parameters, $p_{D}$ = var(deviance)/2, is 1501.6. Rhat is the Gelman-Rubin convergence diagnostic factor \citep{Gelman:1992}.}
 \label{CD4}
\end{table}

Although CD4 count is widely used as a biomarker for treatment progression while studying the efficacy of HIV/AIDS treatments, the CD4 variation and mortality association is not clear yet. Our findings suggest that CD4 cumulative variations significantly impact the hazard, independent of the treatment effect and relevant variables. Targeting CD4 variation may provide an innovative mechanism for new drug development in the future. Furthermore, our findings may help further evaluate CD4 count as a surrogate endpoint instead of overall survival in clinical trials.
\section{Discussion}
\label{s:discuss}
\begin{sidewaysfigure}[ht]
\centering
   \includegraphics[height=20em,width=1\linewidth]{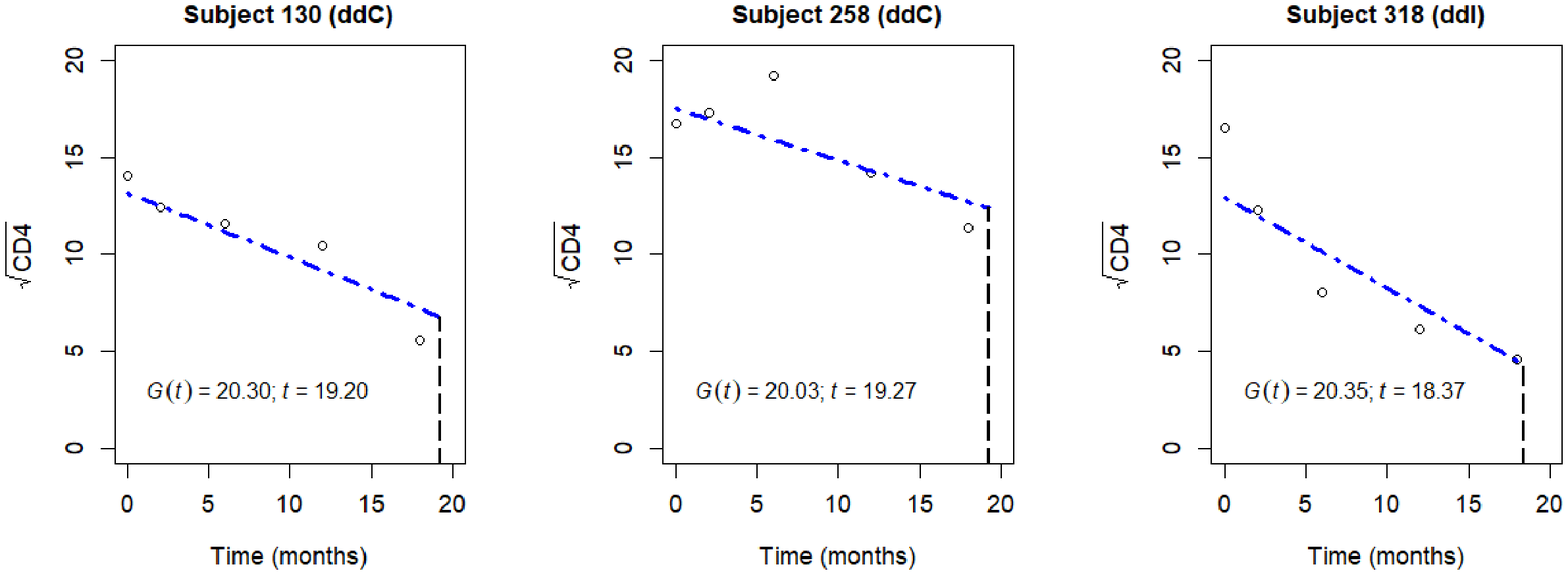}
   \caption{Illustration of high-risk patients in the HIV/AIDS clinical trial. Three patients (ID: 130, 258, 318) are identified as high-risk patients based on the reference ranges of two risk factors: (i) the estimated cumulative variation: $\leq$ the $95^{th}$ percentile; and (ii) the observed survival time: $\geq$ the $95^{th}$ percentile. When a patient fall outside of the limits, it may signal that the individual require further investigation and special care.}
   \label{at_risk}
\end{sidewaysfigure}

The strengths of BALSAM are featured by the innovative quantification and statistical inference of cumulative variations, which can be embedded in the standard joint models with the shared random effects and cumulative effects. Our model can capture helpful information from longitudinal biomarkers to improve hazard estimation in survival analysis. Cumulative variations assessed by arc length may help identify at-risk or high-risk populations, as illustrated in Figure \ref{at_risk}, paving the way for personalized medicine. It can be extended to more types of time-to-event data, including those with competing risks, recurrent events, left censoring, and interval censoring. Our proposed model may be extended to high-dimensional variable selection and latent class procedures. The data-driven time-varying optimal latency window may be explored in BALSAM to accommodate more applications. Besides MCMC algorithms, the Expectation-Maximization (EM) algorithm is often utilized by the MLE approach for joint models. Special care needs to be taken for possible convergence issues 
as the likelihood increases with each iteration \citep{dempster:1977}. 

Our research will have a wide range of applications, including hematology/oncology, cardiology, mental illness, and clinical trials. For instance, it may help identify and compare cancer biomarkers. Adding cumulative variations may help detect abnormal variability of heart rates resulting from diabetes or congestive heart failure. Our methodology may be able to capture mood swings to improve the diagnostic accuracy of mental illness. The signs of elusive diseases, such as Alzheimer's, may be predicted by monitoring and measuring brain activity. With rapid advancements in wearable devices, real-time data for time-varying health measurements are becoming accessible to everyone. Data variations of these measurements may help optimize mobile health interventions' design in chronic disease management, personalized medical support, and health recommendations \citep{Shani:2018,Lee:2018,Bidargaddi:2020}. More exciting applications may include musical composition and speech recognition through arc length to differentiate patterns of sounds and instruments. 

\backmatter

\section*{SUPPORTING INFORMATION}
The computing code and data for the HIV/AIDS clinical trial in Section 4 are available at \url{https://github.com/yangaouic/BALSAM/blob/main/Final_JAGS_CD4_Yan_Gao_26APR2022_GitHub.R}.
\vspace*{-8pt}

\bibliographystyle{biom} \bibliography{biomsample}

\begin{thebibliography}{}

\bibitem[\protect\citeauthoryear{Abrams, Goldman, Launer, Korvick, Neaton,
  Crane, Grodesky, Wakefield, Muth, Kornegay, et~al\mbox{.}}{Abrams
  et~al.}{1994}]{Abrams:1994}
Abrams, D.~I., Goldman, A.~I., Launer, C., Korvick, J.~A., Neaton, J.~D.,
  Crane, L.~R., Grodesky, M., Wakefield, S., Muth, K., Kornegay, S., et~al.
  (1994).
\newblock A comparative trial of didanosine or zalcitabine after treatment with
  zidovudine in patients with human immunodeficiency virus infection.
\newblock {\em New England Journal of Medicine} {\bf 330,} 657--662.

\bibitem[\protect\citeauthoryear{Allison}{Allison}{1978}]{allison:1978}
Allison, P.~D. (1978).
\newblock Measures of inequality.
\newblock {\em American Sociological Review} {\bf 43,} 865--880.

\bibitem[\protect\citeauthoryear{Alsefri, Sudell, García-Fiñana, and
  Kolamunnage-Dona}{Alsefri et~al.}{2020}]{Alsefri:2020}
Alsefri, M., Sudell, M., García-Fiñana, M., and Kolamunnage-Dona, R. (2020).
\newblock Bayesian joint modelling of longitudinal and time to event data: a
  methodological review.
\newblock {\em BMC Medical Research Methodology} {\bf 20,} 94--94.

\bibitem[\protect\citeauthoryear{Anagnostou, Yarchoan, Hansen, Wang, Verde,
  Sharon, Collyar, Chow, and Forde}{Anagnostou et~al.}{2017}]{anagnostou:2017}
Anagnostou, V., Yarchoan, M., Hansen, A.~R., Wang, H., Verde, F., Sharon, E.,
  Collyar, D., Chow, L.~Q., and Forde, P.~M. (2017).
\newblock Immuno-oncology trial endpoints: capturing clinically meaningful
  activity.
\newblock {\em Clinical Cancer Research} {\bf 23,} 4959--4969.

\bibitem[\protect\citeauthoryear{Andrieu and Thoms}{Andrieu and
  Thoms}{2008}]{Andrieu:2008}
Andrieu, C. and Thoms, J. (2008).
\newblock {A tutorial on adaptive MCMC}.
\newblock {\em Statistics and Computing} {\bf 18,} 343--373.

\bibitem[\protect\citeauthoryear{Awoke~Ayele, Worku, Kebede, Zuma, Kasim, and
  Shkedy}{Awoke~Ayele et~al.}{2019}]{Awoke:2019}
Awoke~Ayele, T., Worku, A., Kebede, Y., Zuma, K., Kasim, A., and Shkedy, Z.
  (2019).
\newblock Model-based prediction of {CD4} cells counts in {HIV}-infected adults
  on antiretroviral therapy in {N}orthwest {E}thiopia: A flexible mixed effects
  approach.
\newblock {\em PLOS One} {\bf 14,} e0218514.

\bibitem[\protect\citeauthoryear{Battistini~Garcia and
  Guzman}{Battistini~Garcia and Guzman}{2021}]{Battistini:2021}
Battistini~Garcia, S.~A. and Guzman, N. (2021).
\newblock {\em Acquired Immune Deficiency Syndrome CD4+ Count}.
\newblock StatPearls, Treasure Island, US.

\bibitem[\protect\citeauthoryear{Bender, Groll, and Scheipl}{Bender
  et~al.}{2018}]{Bender:2018}
Bender, A., Groll, A., and Scheipl, F. (2018).
\newblock A generalized additive model approach to time-to-event analysis.
\newblock {\em Statistical Modelling} {\bf 18,} 299--321.

\bibitem[\protect\citeauthoryear{Bender, Scheipl, Hartl, Day, and
  K{\"u}chenhoff}{Bender et~al.}{2019}]{Bender:2019}
Bender, A., Scheipl, F., Hartl, W., Day, A.~G., and K{\"u}chenhoff, H. (2019).
\newblock Penalized estimation of complex, non-linear exposure-lag-response
  associations.
\newblock {\em Biostatistics} {\bf 20,} 315--331.

\bibitem[\protect\citeauthoryear{Bender, Augustin, and Blettner}{Bender
  et~al.}{2005}]{Bender:2005}
Bender, R., Augustin, T., and Blettner, M. (2005).
\newblock Generating survival times to simulate {C}ox proportional hazards
  models.
\newblock {\em Statistics in Medicine} {\bf 24,} 1713--1723.

\bibitem[\protect\citeauthoryear{Berry, Carroll, and Ruppert}{Berry
  et~al.}{2002}]{berry:2002}
Berry, S.~M., Carroll, R.~J., and Ruppert, D. (2002).
\newblock Bayesian smoothing and regression splines for measurement error
  problems.
\newblock {\em Journal of the American Statistical Association} {\bf 97,}
  160--169.

\bibitem[\protect\citeauthoryear{Bidargaddi, Schrader, Klasnja, Licinio, and
  Murphy}{Bidargaddi et~al.}{2020}]{Bidargaddi:2020}
Bidargaddi, N., Schrader, G., Klasnja, P., Licinio, J., and Murphy, S. (2020).
\newblock Designing m-health interventions for precision mental health support.
\newblock {\em Translational Psychiatry} {\bf 10,} 222--222.

\bibitem[\protect\citeauthoryear{Borst, Ahrends, B{\k{a}}ba{\l}a, Melief, and
  Kastenm{\"u}ller}{Borst et~al.}{2018}]{Borst:2018}
Borst, J., Ahrends, T., B{\k{a}}ba{\l}a, N., Melief, C.~J., and
  Kastenm{\"u}ller, W. (2018).
\newblock {CD4}+ {T} cell help in cancer immunology and immunotherapy.
\newblock {\em Nature Reviews Immunology} {\bf 18,} 635--647.

\bibitem[\protect\citeauthoryear{Briggs, Cochran, Gillett, and Schulz}{Briggs
  et~al.}{2018}]{Briggs:2015}
Briggs, W., Cochran, L., Gillett, B., and Schulz, E. (2018).
\newblock {\em Calculus: early transcendentals (3rd Edition)}.
\newblock Pearson, New York, US.

\bibitem[\protect\citeauthoryear{Brown}{Brown}{2009}]{Brown:2009}
Brown, E.~R. (2009).
\newblock Assessing the association between trends in a biomarker and risk of
  event with an application in pediatric {HIV/AIDS}.
\newblock {\em The Annals of Applied Statistics} {\bf 3,} 1163.

\bibitem[\protect\citeauthoryear{Cho{\l}oniewski, Sienkiewicz, Dretnik, Leban,
  Thelwall, and Ho{\l}yst}{Cho{\l}oniewski et~al.}{2020}]{choloniewski:2020}
Cho{\l}oniewski, J., Sienkiewicz, J., Dretnik, N., Leban, G., Thelwall, M., and
  Ho{\l}yst, J.~A. (2020).
\newblock A calibrated measure to compare fluctuations of different entities
  across timescales.
\newblock {\em Scientific Reports} {\bf 10,} 1--16.

\bibitem[\protect\citeauthoryear{Crainiceanu, Ruppert, and Wand}{Crainiceanu
  et~al.}{2005}]{crainiceanu:2005}
Crainiceanu, C., Ruppert, D., and Wand, M.~P. (2005).
\newblock Bayesian analysis for penalized spline regression using {WinBUGS}.
\newblock {\em Journal of Statistical Software} {\bf 14,}.

\bibitem[\protect\citeauthoryear{Dafni and Tsiatis}{Dafni and
  Tsiatis}{1998}]{Dafni:1998}
Dafni, U.~G. and Tsiatis, A.~A. (1998).
\newblock Evaluating surrogate markers of clinical outcome when measured with
  error.
\newblock {\em Biometrics} {\bf 54,} 1445--1462.

\bibitem[\protect\citeauthoryear{De~Gruttola and Tu}{De~Gruttola and
  Tu}{1994}]{De:1994}
De~Gruttola, V. and Tu, X.~M. (1994).
\newblock Modelling progression of {CD4}-lymphocyte count and its relationship
  to survival time.
\newblock {\em Biometrics} {\bf 50,} 1003--1014.

\bibitem[\protect\citeauthoryear{Delgado and Guddati}{Delgado and
  Guddati}{2021}]{delgado:2021}
Delgado, A. and Guddati, A.~K. (2021).
\newblock Clinical endpoints in oncology - a primer.
\newblock {\em American Journal of Cancer Research} {\bf 11,} 1121.

\bibitem[\protect\citeauthoryear{Dempster, Laird, and Rubin}{Dempster
  et~al.}{1977}]{dempster:1977}
Dempster, A.~P., Laird, N.~M., and Rubin, D.~B. (1977).
\newblock Maximum likelihood from incomplete data via the {EM} algorithm.
\newblock {\em Journal of the Royal Statistical Society, Series B: Methodology}
  {\bf 39,} 1--22.

\bibitem[\protect\citeauthoryear{DeRogatis, Viramontes, Neubert, and
  Tinoco}{DeRogatis et~al.}{2021}]{derogatis:2021}
DeRogatis, J.~M., Viramontes, K.~M., Neubert, E.~N., and Tinoco, R. (2021).
\newblock {PSGL}-1 immune checkpoint inhibition for {CD4+ T} cell cancer
  immunotherapy.
\newblock {\em Frontiers in Immunology} {\bf 12,} 15.

\bibitem[\protect\citeauthoryear{Desmée, Mentré, Veyrat-Follet, and
  Guedj}{Desmée et~al.}{2015}]{Solene:2015}
Desmée, S., Mentré, F., Veyrat-Follet, C., and Guedj, J. (2015).
\newblock Nonlinear mixed-effect models for prostate-specific antigen kinetics
  and link with survival in the context of metastatic prostate cancer: a
  comparison by simulation of two-stage and joint approaches.
\newblock {\em The AAPS Journal} {\bf 17,} 691--699.

\bibitem[\protect\citeauthoryear{Desmée, Mentré, Veyrat-Follet, Sébastien,
  and Guedj}{Desmée et~al.}{2017}]{Solene:2017}
Desmée, S., Mentré, F., Veyrat-Follet, C., Sébastien, B., and Guedj, J.
  (2017).
\newblock Nonlinear joint models for individual dynamic prediction of risk of
  death using {Hamiltonian Monte Carlo}: application to metastatic prostate
  cancer.
\newblock {\em BMC Medical Research Methodology} {\bf 17,} 105--105.

\bibitem[\protect\citeauthoryear{Duane, Kennedy, Pendleton, and Roweth}{Duane
  et~al.}{1987}]{Duane:1987}
Duane, S., Kennedy, A., Pendleton, B.~J., and Roweth, D. (1987).
\newblock {Hybrid Monte Carlo}.
\newblock {\em Physics Letters B} {\bf 195,} 216--222.

\bibitem[\protect\citeauthoryear{Fano}{Fano}{1947}]{fano:1947}
Fano, U. (1947).
\newblock Ionization yield of radiations. {II.} the fluctuations of the number
  of ions.
\newblock {\em Physical Review} {\bf 72,} 26.

\bibitem[\protect\citeauthoryear{Farkona, Diamandis, and Blasutig}{Farkona
  et~al.}{2016}]{farkona:2016}
Farkona, S., Diamandis, E.~P., and Blasutig, I.~M. (2016).
\newblock Cancer immunotherapy: the beginning of the end of cancer?
\newblock {\em BMC Medicine} {\bf 14,} 1--18.

\bibitem[\protect\citeauthoryear{Faucett and Thomas}{Faucett and
  Thomas}{1996}]{FAUCETT:1996}
Faucett, C.~L. and Thomas, D.~C. (1996).
\newblock Simultaneously modelling censored survival data and repeatedly
  measured covariates: A {G}ibbs sampling approach.
\newblock {\em Statistics in Medicine} {\bf 15,} 1663--1685.

\bibitem[\protect\citeauthoryear{Feigin}{Feigin}{2004}]{feigin:2004}
Feigin, A. (2004).
\newblock Evidence from biomarkers and surrogate endpoints.
\newblock {\em NeuroRx} {\bf 1,} 323--330.

\bibitem[\protect\citeauthoryear{Fiore and D’Avolio}{Fiore and
  D’Avolio}{2011}]{Fiore:2011}
Fiore, L.~D. and D’Avolio, L.~W. (2011).
\newblock {Detours on the Road to Personalized Medicine: Barriers to Biomarker
  Validation and Implementation}.
\newblock {\em JAMA} {\bf 306,} 1914--1915.

\bibitem[\protect\citeauthoryear{Fleming and Powers}{Fleming and
  Powers}{2012}]{fleming:2012}
Fleming, T.~R. and Powers, J.~H. (2012).
\newblock Biomarkers and surrogate endpoints in clinical trials.
\newblock {\em Statistics in Medicine} {\bf 31,} 2973--2984.

\bibitem[\protect\citeauthoryear{Gasparrini}{Gasparrini}{2014}]{Gasparrini:2014}
Gasparrini, A. (2014).
\newblock Modeling exposure–lag–response associations with distributed lag
  non-linear models.
\newblock {\em Statistics in Medicine} {\bf 33,} 881--899.

\bibitem[\protect\citeauthoryear{Gasparrini, Armstrong, and Kenward}{Gasparrini
  et~al.}{2010}]{Gasparrini:2010}
Gasparrini, A., Armstrong, B., and Kenward, M.~G. (2010).
\newblock Distributed lag non-linear models.
\newblock {\em Statistics in Medicine} {\bf 29,} 2224--2234.

\bibitem[\protect\citeauthoryear{Gasparrini, Scheipl, Armstrong, and
  Kenward}{Gasparrini et~al.}{2017}]{Gasparrini:2017}
Gasparrini, A., Scheipl, F., Armstrong, B., and Kenward, M.~G. (2017).
\newblock A penalized framework for distributed lag non-linear models.
\newblock {\em Biometrics} {\bf 73,} 938--948.

\bibitem[\protect\citeauthoryear{Gelfand and Smith}{Gelfand and
  Smith}{1990}]{gelfand:1990}
Gelfand, A.~E. and Smith, A.~F. (1990).
\newblock Sampling-based approaches to calculating marginal densities.
\newblock {\em Journal of the American Statistical Association} {\bf 85,}
  398--409.

\bibitem[\protect\citeauthoryear{Gelman and Rubin}{Gelman and
  Rubin}{1992}]{Gelman:1992}
Gelman, A. and Rubin, D.~B. (1992).
\newblock Inference from iterative simulation using multiple sequences.
\newblock {\em Statistical Science} {\bf 7,} 457--472.

\bibitem[\protect\citeauthoryear{González-Navajas, Elkord, and
  Lee}{González-Navajas et~al.}{2021}]{Gonza:2021}
González-Navajas, J.~M., Elkord, E., and Lee, J. (2021).
\newblock Editorial: {CD4 + T} cells in cancer immunotherapies.
\newblock {\em Frontiers in Immunology} {\bf 12,} 737615--737615.

\bibitem[\protect\citeauthoryear{Guedj}{Guedj}{2018}]{Guedj:2018}
Guedj, J. (2018).
\newblock Joint modeling of tumor kinetic and overall survival.
\newblock https://www.fda.gov/media/113417/download.

\bibitem[\protect\citeauthoryear{Henderson}{Henderson}{2002}]{Henderson:2002}
Henderson, R. (2002).
\newblock Identification and efficacy of longitudinal markers for survival.
\newblock {\em Biostatistics} {\bf 3,} 33--50.

\bibitem[\protect\citeauthoryear{Hoffman and Gelman}{Hoffman and
  Gelman}{2014}]{Hoffman:2014}
Hoffman, M.~D. and Gelman, A. (2014).
\newblock The {No-U-Turn} sampler: Adaptively setting path lengths in
  {Hamiltonian Monte Carlo}.
\newblock {\em Journal of Machine Learning Research} {\bf 15,} 1593--1623.

\bibitem[\protect\citeauthoryear{Ibrahim, Chen, and Sinha}{Ibrahim
  et~al.}{2001}]{Ibrahim:2001}
Ibrahim, J.~G., Chen, M., and Sinha, D. (2001).
\newblock {\em Bayesian survival analysis}.
\newblock Springer series in statistics. Springer, New York, US.

\bibitem[\protect\citeauthoryear{Ibrahim, Chu, and Chen}{Ibrahim
  et~al.}{2010}]{Ibrahim:2010}
Ibrahim, J.~G., Chu, H., and Chen, L.~M. (2010).
\newblock Basic concepts and methods for joint models of longitudinal and
  survival data.
\newblock {\em Journal of Clinical Oncology} {\bf 28,} 2796--2801.

\bibitem[\protect\citeauthoryear{Lee, Choi, Lee, and Jiang}{Lee
  et~al.}{2018}]{Lee:2018}
Lee, J.-A., Choi, M., Lee, S.~A., and Jiang, N. (2018).
\newblock Effective behavioral intervention strategies using mobile health
  applications for chronic disease management: a systematic review.
\newblock {\em BMC Medical Informatics and Decision Making} {\bf 18,} 12--12.

\bibitem[\protect\citeauthoryear{Lin, Fischl, and Schoenfeld}{Lin
  et~al.}{1993}]{lin:1993}
Lin, D., Fischl, M.~A., and Schoenfeld, D. (1993).
\newblock Evaluating the role of {CD4}-lymphocyte counts as surrogate endpoints
  in human immunodeficiency virus clinical trials.
\newblock {\em Statistics in Medicine} {\bf 12,} 835--842.

\bibitem[\protect\citeauthoryear{MacKay}{MacKay}{2003}]{MacKay:2003}
MacKay, D. J.~C. (2003).
\newblock {\em Information theory, inference, and learning algorithms}.
\newblock Cambridge University Press, Cambridge, UK.

\bibitem[\protect\citeauthoryear{McCullagh and Nelder}{McCullagh and
  Nelder}{1989}]{Mccullagh:1989}
McCullagh, P. and Nelder, A.~J. (1989).
\newblock {\em Generalized linear models (2nd edition)}.
\newblock Chapman $\&$ Hall/CRC, London, UK.

\bibitem[\protect\citeauthoryear{McCulloch}{McCulloch}{2003}]{McCulloch:2003}
McCulloch, C.~E. (2003).
\newblock {\em Chapter 4: Generalized linear mixed models ({GLMMs})}, volume~7.
\newblock Institute of Mathematical Statistics, US.

\bibitem[\protect\citeauthoryear{Nahum-Shani, Smith, Spring, Collins,
  Witkiewitz, Tewari, and Murphy}{Nahum-Shani et~al.}{2018}]{Shani:2018}
Nahum-Shani, I., Smith, S.~N., Spring, B.~J., Collins, L.~M., Witkiewitz, K.,
  Tewari, A., and Murphy, S.~A. (2018).
\newblock Just-in-time adaptive interventions {(JITAIs)} in mobile health: Key
  components and design principles for ongoing health behavior support.
\newblock {\em Annals of Behavioral Medicine} {\bf 52,} 446--462.

\bibitem[\protect\citeauthoryear{Neal}{Neal}{1993}]{Neal:1993}
Neal, R.~M. (1993).
\newblock {\em Probabilistic inference using {Markov Chain Monte Carlo
  methods}}.
\newblock University of Toronto, Toronto, CAN.

\bibitem[\protect\citeauthoryear{Neal}{Neal}{1996}]{Neal:1996}
Neal, R.~M. (1996).
\newblock {\em Bayesian learning for neural networks}.
\newblock Springer, New York, US.

\bibitem[\protect\citeauthoryear{Nesterov}{Nesterov}{2009}]{Nesterov:2009}
Nesterov, Y. (2009).
\newblock Primal-dual subgradient methods for convex problems.
\newblock {\em Mathematical Programming} {\bf 120,} 221--259.

\bibitem[\protect\citeauthoryear{Pawitan and Self}{Pawitan and
  Self}{1993}]{Pawitan:1993}
Pawitan, Y. and Self, S. (1993).
\newblock Modeling disease market processes in aids.
\newblock {\em Journal of the American Statistical Association} {\bf 88,}
  719--726.

\bibitem[\protect\citeauthoryear{Redfield and Burke}{Redfield and
  Burke}{1988}]{Redfield:1988}
Redfield, R.~R. and Burke, D.~S. (1988).
\newblock {HIV} infection: The clinical picture.
\newblock {\em Scientific American} {\bf 259,} 90--99.

\bibitem[\protect\citeauthoryear{Reeves and Derdeyn}{Reeves and
  Derdeyn}{2007}]{reeves:2007}
Reeves, J.~D. and Derdeyn, C.~A. (2007).
\newblock {\em Entry inhibitors in HIV therapy}.
\newblock Springer, New York, US.

\bibitem[\protect\citeauthoryear{Robb, McInnes, and Califf}{Robb
  et~al.}{2016}]{robb:2016}
Robb, M.~A., McInnes, P.~M., and Califf, R.~M. (2016).
\newblock Biomarkers and surrogate endpoints: developing common terminology and
  definitions.
\newblock {\em JAMA} {\bf 315,} 1107--1108.

\bibitem[\protect\citeauthoryear{Robert, Elvira, Tawn, and Wu}{Robert
  et~al.}{2018}]{Robert:2018}
Robert, P.~C., Elvira, V., Tawn, N., and Wu, C. (2018).
\newblock Accelerating {MCMC} algorithms.
\newblock {\em WIREs Computational Statistics} {\bf 10,} e1435.

\bibitem[\protect\citeauthoryear{Ruppert, Wand, and Carroll}{Ruppert
  et~al.}{2003}]{ruppert:2003}
Ruppert, D., Wand, M.~P., and Carroll, R.~J. (2003).
\newblock {\em Semiparametric regression}.
\newblock Number~12 in Statistical and Probabilistic Mathematics. Cambridge
  University Press.

\bibitem[\protect\citeauthoryear{Self and Pawitan}{Self and
  Pawitan}{1992}]{Self:1992}
Self, S. and Pawitan, Y. (1992).
\newblock Modeling a marker of disease progression and onset of disease.
\newblock In Jewell, N.~P., Dietz, K., and Farewell, V.~T., editors, {\em AIDS
  Epidemiology: Methodological Issues}, pages 231--255. Birkh{\"a}user Boston,
  Boston, MA.

\bibitem[\protect\citeauthoryear{Strimbu and Tavel}{Strimbu and
  Tavel}{2010}]{strimbu:2010}
Strimbu, K. and Tavel, J.~A. (2010).
\newblock What are biomarkers?
\newblock {\em Current Opinion in HIV and AIDS} {\bf 5,} 463.

\bibitem[\protect\citeauthoryear{Sylvestre and Abrahamowicz}{Sylvestre and
  Abrahamowicz}{2009}]{Sylvestre:2009}
Sylvestre, M.-P. and Abrahamowicz, M. (2009).
\newblock Flexible modeling of the cumulative effects of time-dependent
  exposures on the hazard.
\newblock {\em Statistics in Medicine} {\bf 28,} 3437--3453.

\bibitem[\protect\citeauthoryear{Tsiatis, Degruttola, and Wulfsohn}{Tsiatis
  et~al.}{1995}]{Tsiatis:1995}
Tsiatis, A.~A., Degruttola, V., and Wulfsohn, M.~S. (1995).
\newblock Modeling the relationship of survival to longitudinal data measured
  with error. applications to survival and {CD4} counts in patients with aids.
\newblock {\em Journal of the American Statistical Association} {\bf 90,}
  27--37.

\bibitem[\protect\citeauthoryear{{U.S. Food and Drug Administration}}{{U.S.
  Food and Drug Administration}}{2021}]{FDA:2021}
{U.S. Food and Drug Administration} (2021).
\newblock {FDA} facts: Biomarkers and surrogate endpoints.
\newblock
  https://www.fda.gov/about-fda/innovation-fda/fda-facts-biomarkers-and-surrogate-endpoints.

\bibitem[\protect\citeauthoryear{Wickramarachchi, Gallagher, and
  Lund}{Wickramarachchi et~al.}{2014}]{Wick:2014}
Wickramarachchi, T.~D., Gallagher, C., and Lund, R. (2014).
\newblock Arc length asymptotics for multivariate time series.
\newblock {\em Applied Stochastic Models in Business and Industry} {\bf 31,}
  264--281.

\bibitem[\protect\citeauthoryear{Wulfsohn and Tsiatis}{Wulfsohn and
  Tsiatis}{1997}]{Wulfsohn:1997}
Wulfsohn, M.~S. and Tsiatis, A.~A. (1997).
\newblock A joint model for survival and longitudinal data measured with error.
\newblock {\em Biometrics} {\bf 53,} 330--339.

\bibitem[\protect\citeauthoryear{Zanetti}{Zanetti}{2015}]{zanetti:2015}
Zanetti, M. (2015).
\newblock Tapping {CD4 T} cells for cancer immunotherapy: the choice of
  personalized genomics.
\newblock {\em The Journal of Immunology} {\bf 194,} 2049--2056.

\bibitem[\protect\citeauthoryear{Zhang, Pilar, Wang, Liu, Pang, Brownson,
  Colditz, Liang, and He}{Zhang et~al.}{2020}]{zhang:2020}
Zhang, J., Pilar, M.~R., Wang, X., Liu, J., Pang, H., Brownson, R.~C., Colditz,
  G.~A., Liang, W., and He, J. (2020).
\newblock Endpoint surrogacy in oncology {P}hase 3 randomised controlled
  trials.
\newblock {\em British Journal of Cancer} {\bf 123,} 333--334.

\end{thebibliography}


\appendix


\section{}
\subsection{Observed Joint Likelihood for Model I }
The observed joint likelihood $f(t_{i},\delta_{i}, \bm{z}_{i}|\bm{\theta}) $ for the $i$th subject in Model I is given by
\begin{align*}
    \bigintss \bigg\{ &
    \lambda^{\delta_{i}} \; \exp \bigg\{ \bigg( x_{i1}\beta + \alpha t_{i} \sqrt{1+b_{i1}^{2}} \bigg) \delta_{i} \bigg \}  \\
    & \;  \times \; \exp\bigg\{ -\lambda \exp\{x_{i1}\beta\} \bigg(\frac{1} {\alpha \sqrt{1+ b_{i1}^{2}}}\bigg) \bigg(\exp \bigg \{\alpha t_{i}  \sqrt{1+ b_{i1}^{2}}\bigg \}-1\bigg) \bigg \} \\
   &  \times \prod_{j=1}^{n_{i}} \bigg \{ \frac{1} { \sigma \sqrt{2\pi} } \exp \bigg \{-\frac{(z_{ij}-b_{i0} - b_{i1}s_{ij})^{2}} {2\sigma^{2}} \bigg \} \bigg \} \\
   &  \times 2 (\pi)^{-\frac{K}{2}} \det({\bm{\Sigma}})^{-\frac{1}{2}} \exp \Bigg\{-\frac{1}{2} ({\bm{b}_{i}} - \bm{\mu})^{\mathrm{T}} {\bm{\Sigma}^{-1}} ({\bm{b}_{i} - \bm{\mu}) \Bigg \} \Bigg \} \;  \mathrm{d} {\bm{b}_{i}} }
\end{align*}
\subsection{Observed Joint Likelihood for Model II}
The observed joint likelihood $f(t_{i},\delta_{i}, \bm{z}_{i}|\bm{\theta}) $ for the $i$th subject in Model II is given by
\begin{align*}
    \bigints \Bigg\{ &
    \lambda^{\delta_{i}} \exp \Bigg\{ \Bigg( x_{i1}\beta + \alpha \bigints_{s=0}^{t_{i}} \sqrt{1+ \bigg \{ \sum_{l=1}^{K} b_{il}B'_{l}(s)\bigg \}^{2} } \; \mathrm{d}s \Bigg) \delta_{i} \Bigg \}  \\
    &  \times \exp\Bigg\{ - \bigints_{s=0}^{t_{i}} \Bigg\{ \lambda \exp\Bigg\{ x_{i1}\beta + \alpha \bigints_{u=0}^{s} \sqrt{1+ \bigg \{ \sum_{l=1}^{K} b_{il}B'_{l}(u) \bigg \}^{2} } \; \mathrm{d}u  \Bigg \} \Bigg \} \; \mathrm{d} s  \Bigg \} \\
   & \times \prod_{j=1}^{n_{i}} \Bigg \{ \frac{1}{ \sigma \sqrt{2\pi} } \exp\Bigg \{-
   \frac{1}{2\sigma^{2}}
   \bigg ( z_{ij}- \displaystyle \sum_{l=1}^{K} b_{il}B_{l}(s_{ij}) \bigg )^{2}  \Bigg \} \Bigg \} \\
   &  \times  (2\pi)^{-\frac{K}{2}} \det({\bm{\Sigma}})^{-\frac{1}{2}} \exp \Bigg\{-\frac{1}{2} ({\bm{b}_{i}}  - \bm{\mu})^{\mathrm{T}} {\bm{\Sigma}^{-1}} ({\bm{b}_{i}}  - \bm{\mu}) \Bigg \} \Bigg \} \;  \mathrm{d} {\bm{b}_{i}} 
\end{align*}
\subsection{Computational Challenge}
We write our model based on the observed likelihood by integrating out the random effects to reduce the total number of parameters in the MCMC algorithms. This way aims to speed up the posterior estimation by marginalizing over random effects in the likelihood calculation. However, we have to treat a random effect as a parameter in the Bayesian software implementation, resulting in a significant increase in the dimension of parameter space. We conjecture that the posterior estimation of the parameters of interest should be equivalent because the random effects are marginalized eventually in either way. BALSAM revolves around two nested integrals in the cumulative hazard; that is, the upper bound of the inner integral is the variable of the outer in (\ref{eqn_H_comp}). Notably, the inner integral rests on the power of the exponential function. Thus, it is not able to integrate over a region as the classical integration often does for a double or iterated integral. The computation may become complicated when the closed-form solution is not feasible.
\begin{equation} \label{eqn_H_comp}
    H(t) = \underbrace{ \int_{s=0}^{t}  \bigg \{ \lambda_{0}(s) \exp \Big \{  \bm{x}'\bm{\beta} +  \alpha  \overbrace{\int_{u=0}^{s} |g'(u)|\mathrm{d}u }^{\mathrm{inner}} \Big \} \bigg \}  \;  \mathrm{d} s}_{\mathrm{outer}}
\end{equation}

The computational challenge results from the sequential order. The ordinary numerical approximation to the integral results in extremely slow convergence if the outer integral has to wait until the inner integral is completed for each time point. Furthermore, the random walk MCMC algorithms require sequential iterations while updating parameters. The likelihood calculation needs to search over all subjects. These sequential iterations require an innovative algorithm strategy. Thus, we adopt a series of efficient algorithms to compute those integrals and accelerate the successful construction of Markov chains, including matrix vectorization, Cholesky decomposition of the covariance matrix, and accelerating MCMC methods. 
\label{lastpage}

\end{document}